\documentclass[aps,notitlepage,nofootinbib,longbibliography,twocolumn,superscriptaddress,10pt]{revtex4-2}

\usepackage{bbold}
\usepackage{simplewick}

\usepackage[usenames,dvipsnames]{color}
\usepackage[colorlinks=true,citecolor=PineGreen,linkcolor=blue]{hyperref}

\usepackage{amsmath}
\usepackage{amsthm, amssymb}
\usepackage{physics}
\usepackage{enumitem}
\usepackage{slashed}
\usepackage{graphicx}
\usepackage{xcolor}
\usepackage{tabularx}
\usepackage{tikz}
\usetikzlibrary{calc,fadings,decorations.pathreplacing,shapes,shapes.multipart,arrows,shapes.misc,intersections,positioning,patterns}
\usepackage{bm}
\usepackage{bbm}
\usepackage[colorlinks=true,citecolor=PineGreen,linkcolor=blue]{hyperref}
\usepackage{dsfont}
\usepackage{changepage}
\usepackage{array}
\usepackage{soul}
\usepackage[capitalise]{cleveref}
\crefname{section}{Sec.}{Secs.}
\Crefname{section}{Sec.}{Secs.}
\usepackage{xifthen}
\usepackage{xargs}
\usepackage{subcaption}
\usepackage{braket}
\definecolor{ltgreen}{rgb}{0.1,.59,.43}

\begin{document}

\title{The X-Cube Floquet Code}
\author{Zhehao Zhang} 
\affiliation{Department of Physics, 
University of California, Santa Barbara, CA 93106, USA} 
\author{David Aasen}
\affiliation{Microsoft Station Q, Santa Barbara, California 93106-6105 USA} 
\author{Sagar Vijay}
\affiliation{Department of Physics, 
University of California, Santa Barbara, CA 93106, USA} 
\begin{abstract}
Inspired by coupled-layer constructions of fracton orders, we introduce the X-Cube Floquet code, a dynamical quantum error-correcting code where the number of encoded logical qubits grows with system size. The X-Cube Floquet code is defined on a three-dimensional lattice, built from intersecting two-dimensional layers in the $xy$, $yz$, and $xz$ directions, and consists of a periodic sequence of two-qubit measurements which couple the layers together. 
Within a single Floquet cycle, the codespace switches between that of the X-Cube fracton order and layers of entangled, two-dimensional toric codes. The encoded logical qubits' dynamics are analyzed, and we argue that the new code has a non-zero error threshold. We provide a new Hamiltonian realization of the X-Cube model and, more generally, explore the phase diagram related to the sequence of measurements that define the X-Cube Floquet code.
\end{abstract}
\maketitle
\tableofcontents
%


\section{Introduction}
The honeycomb code introduced by Hastings and Haah~\cite{hastings2021dynamically} describes a new class of fault-tolerant quantum error correcting codes, which are defined by a periodic schedule of few-body measurements that lead to the protection of quantum information.  These \emph{Floquet codes} are distinct from stabilizer and subsystem codes \cite{poulin2005stabilizer} since the encoding of logical information can be \emph{dynamical}, evolving during the periodic measurement schedule.  
Apart from harboring dynamical logical qubits, such as in the honeycomb code, the physical qubits in Floquet codes can also exhibit non-trivial dynamics, see Refs.~\cite{Aasen2022,Davydova2022}. 

Floquet codes have a number of conceptually novel properties. One feature of Floquet codes is that they typically use only low-weight measurements (e.g., two-qubit measurements). This property has led to desirable error thresholds for the Hastings-Haah honeycomb code, along with proposed implementations~\cite{Gidney2021,Paetznick2022,Gidney2022b,Haah2022,Wootton2022} which suggest that the honeycomb code is potentially well-suited in near-term platforms for quantum computation.  In addition, certain Floquet codes can implement a non-trivial automorphism of the logical Hilbert space in each measurement period.  In Floquet codes such as the honeycomb code, this transformation is intimately related to automorphisms of the anyons in a topological order, and adiabatic paths of gapped Hamiltonians, a connection which was further explored in Ref.~\cite{Aasen2022}. This automorphism is not only conceptually interesting, but fundamentally important, since the two-qubit operators which are measured in a Floquet period of the honeycomb code cannot be used to define a static encoding of logical information \cite{suchara2011constructions}.  Because of these desirable properties, it is of interest to construct other Floquet codes, particularly ones which can ($i$) protect a larger amount of logical information, and ($ii$) realize other kinds of long-range-entangled states of quantum matter which are potentially more challenging to realize in equilibrium.

In this work, we 
introduce a Floquet code for which the number of logical qubits grows sub-extensively in system size, and which realizes a fracton order at regular timesteps.  Gapped fracton orders are long-range-entangled phases of quantum matter in (3+1)D~\cite{HaahCode,Vijay2015,Vijay2016} which host gapped excitations that are topological in nature and exhibit exotic mobility constraints; as a consequence, gapped fracton orders also exhibit a stable ground state degeneracy that generally grows in system size.
The X-cube model \cite{Vijay2016} provides a well-studied example of a fracton order, whose spectrum contains point-like excitations restricted to move along planes (planons) or lines (lineons), as well as excitations that are completely immobile (fractons). As originally formulated, the X-cube model arises from an exactly solvable stabilizer Hamiltonian which consists of four-qubit and twelve-qubit interactions.

Motivated by constructions of the X-Cube model by coupling together layers of toric codes \cite{vijay2017isotropic,ma2017fracton}, and by recent progress in understanding dynamical quantum codes, we introduce the X-Cube Floquet code, which is defined by a six-step sequence of two-qubit measurements which are repeated.
As the name suggests, an effective X-Cube fracton order materializes once per Floquet period. 
At subsequent points along the measurement path, the system realizes weakly-entangled toric code layers, along with other orders which we argue are foliation equivalent \cite{Shirley2018} to the X-cube fracton order. 
Thus this work brings the X-Cube code one step closer to potential realization by reducing the weight four and twelve stabilizers, to a path of weight-two measurements.    

Our new Floquet code is inspired by the coupled-layer construction of the X-cube model presented in Refs.~\cite{vijay2017isotropic,ma2017fracton}.
For each layer of the construction, we utilize the honeycomb code on the 4.8.8 lattice~\cite{Paetznick2022}.
We couple the layers together by an inter-layer measurement once per measurement period.
The rank of the ISG stabilizes is extensive and encodes a sub-extensively large codespace, which is exponential in the linear dimension of the system.
We argue the resulting code has a non-zero error threshold.
Moreover, the check operators alone define a trivial subsystem code which cannot encode any logical information. We note that the lower-weight measurements, along with the distinct quantum orders realized during each Floquet period distinguish our code from a recent proposal \cite{Davydova2022} for a Floquet code that also realizes fracton order.

We also identify and investigate a closely related 3D lattice Hamiltonian given by summing together the checks used to generate the Floquet code. By tuning the parameters of that Hamiltonian, we could identify phases that are related to the ISG of the X-cube model and decoupled-layers of Toric code. 

The paper is organized as follows. In Sec.~\ref{sec:XCubeFC} we define the X-Cube Floquet code as a particular sequence of two-qubit measurements and we analyze the dynamics using the ISG formalism.
In Sec.~\ref{sec:errorcorrection} we argue the X-Cube Floquet code has a non-zero error threshold.


\section{The X-Cube Floquet code}\label{sec:XCubeFC}

{\bf \emph{ 4.8.8 Code:}} 
The fracton Floquet code that we introduce is based on the two-dimensional 4.8.8 Floquet code \cite{Paetznick2022} and the coupled-layer construction of fracton phases introduced in Ref. \cite{vijay2017isotropic,
ma2017fracton}.  Here, we review the check operators of the  4.8.8 code, which we will use subsequently in the construction of the X-Cube Floquet code.

\begin{figure}[t]
$\begin{array}{cc} \includegraphics[width=34mm]{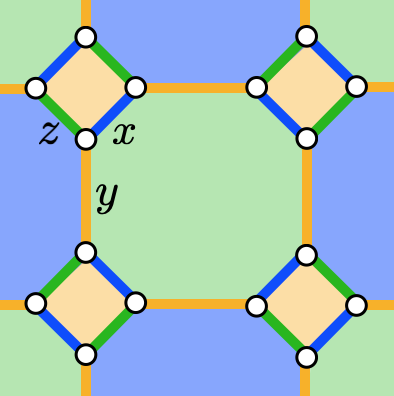} & \,\,
\includegraphics[width=43mm]{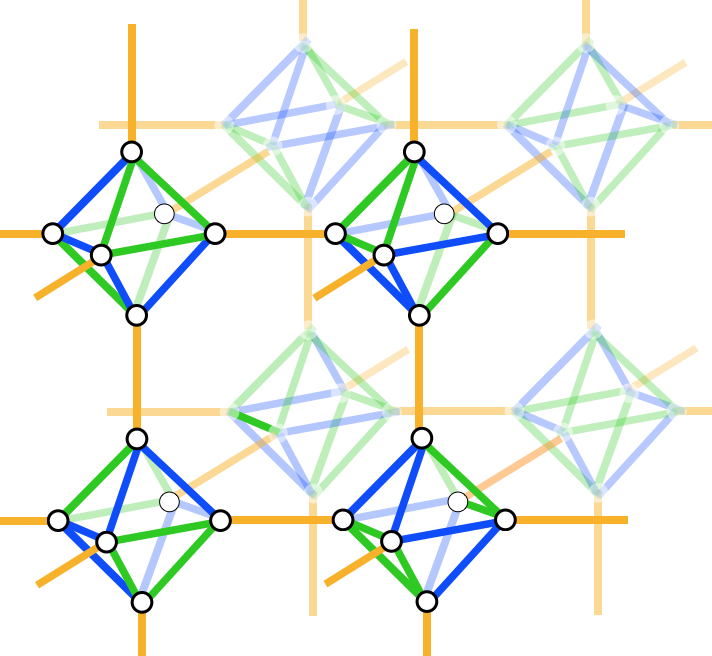}\\
\text{(a)} & \text{(b)}
\end{array}$
     \caption{(a) Underlying lattice for the 4.8.8 Code. Each white vertex contains one qubit. The two-qubit check operators are defined on horizontal and vertical edges as $YY$, on diagonal southwest to northeast edges as $ZZ$, and on diagonal southeast to northwest edges as $XX$. (b) A stack of 4.8.8 lattice in the $xy$, $yz$, and $xz$ directions, where pairs of orthogonal planes overlap along $y$-type edges. Each white vertex in this three-dimensional array now contains two qubits. The color and type of an edge are determined by its location on the supporting 2D 4.8.8 lattice. }
    \label{fig:488}
\end{figure}

The 4.8.8. code is defined on the lattice shown in Fig.\ref{fig:488}a, which hosts a single qubit at each lattice site. Edges are labeled $x$, $y$, or $z$ based on {orientation}. All vertical and horizontal edges are of a $y$-type.  The remaining diagonal bonds are either $x$ or $z$-type as indicated. We define two-qubit ``check operators" acting on each edge based on this labeling. For example, along an edge of type $x$ connecting sites $i$ and $j$, we associate a check operator $X_{i}X_{j}$. We may similarly define two-qubit check operators on $y$- and $z$-type edges, given by the product of Pauli-$Y$ and $Z$ operators, respectively.  
 With this definition of check operators, we note that \emph{plaquette stabilizers}, defined by the product of check operators around elementary square or octagonal plaquettes, manifestly commute with each of the check operators.
 
 In addition to labeling edges based on orientation, we also introduce a coloring of edges (blue, green, or yellow) as shown in Fig. \ref{fig:488}a, which will be useful in defining the X-Cube Floquet code. We first notice that the 4.8.8 lattice is three-colorable, so we can color each plaquette by one of yellow, blue, and green such that two neighboring plaquettes have different colors. We use the coloring convention that the square plaquettes are always yellow, while the octagonal plaquettes alternate in color as blue or green. Each edge is labeled by the same color of the two plaquettes it connect.
 We note that with this convention, the $x$ and $z$-type bonds alternate as blue or green in color, while the $y$-type bond is always yellow. 

{\bf \emph{Check Operators of the X-Cube Floquet Code:}} Inspired by the isotropic layer construction of the X-Cube model~\cite{vijay2017isotropic,ma2017fracton} and the 4.8.8 Floquet code \cite{Paetznick2022}, we now introduce a Floquet code using coupled layers of 4.8.8 lattice. We place $L$ copies of 4.8.8 lattice in the $xy$, $yz$, and $xz$ planes, respectively, so that pairs of mutually orthogonal planes overlap along $y$ edges of the 2D lattice. The 3D lattice contains two qubits at each lattice site. We also require two octagon plaquettes facing each other have the same color. With the above requirements, there can be two different 3D lattices, which differ in  the coloring of the octagonal faces of the elementary cubes. In this paper, we work with the choice of 3D lattice which contains cubes which have either ($i$) all blue faces or ($ii$) four green and two blue faces. The resulting three-dimensional lattice is shown in Fig. \ref{fig:488}(b). 

\begin{figure}[t]
    $\begin{array}{cc}
    \includegraphics[width=37mm]{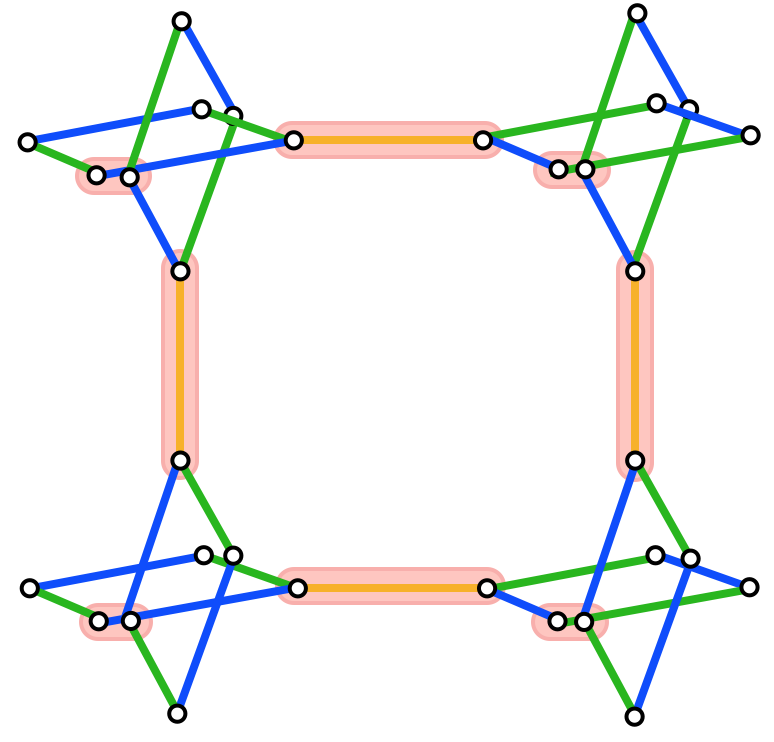} &
    \includegraphics[width=37mm]{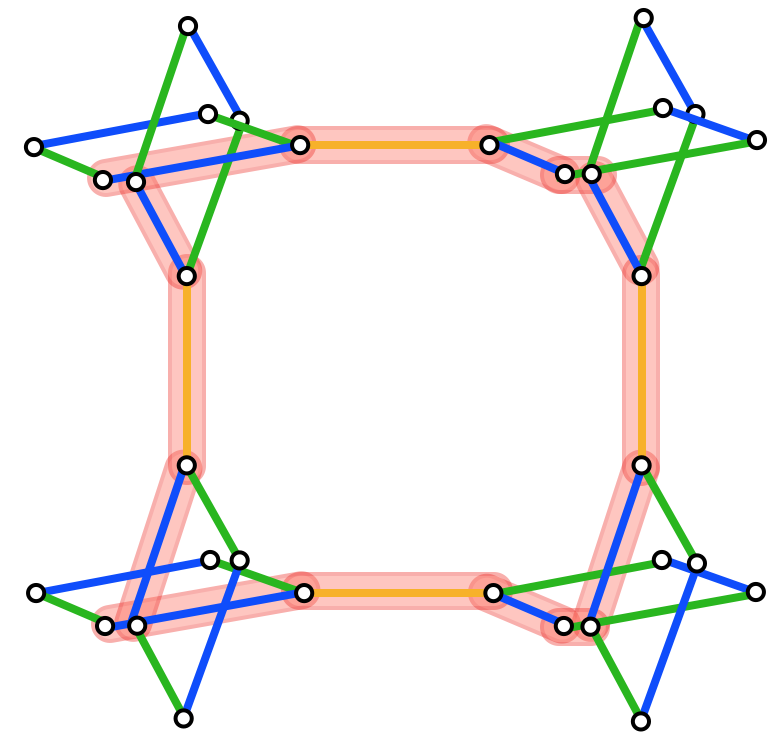}\\
    \text{(a)} & \text{(b)}
    \end{array}$
    \caption{The ``zigzag" stabilizers appear in the ISG in steps 3 and 4 of the Floquet measurement cycle.  The support of this stabilizer at step 3 is shown  (a), by the product of four on-site checks and four yellow checks, which are highlighted in red. In step 4, this stabilizer is given by the product of four on-site checks, four yellow checks, and eight blue checks, which are highlighted in red in (b).}
    \label{fig2}
\end{figure}

We now define check operators that act within a single 4.8.8 layer and act between layers. The intra-layer check operators are the same as the check operators defined in the 4.8.8 Floquet code; we refer to these as green, blue, or yellow check operators, depending on the color of the edge on which the check operator acts. The inter-layer check operators are defined as the product of the $Y$ Pauli operators on the two qubits on the same site.  We refer to these as  on-site checks. 

Importantly, these check operators alone define a trivial subsystem code which does not encode any logical information.  To see this, we consider an $L\times L \times L$ system with periodic boundary condition.  In Appendix \ref{AC}, we determine the gauge group generated by all check operators has dimension $20L^3 - 3L$. The stabilizer group is generated by plaquette stabilizers and stabilizers supported on homologically nontrivial cycles. We find the stabilizer group has dimension $4L^3 + 3L$ (See Appendix \ref{AC}). Hence, there are $g = ((20L^3 - 3L) - (4L^3 + 3L))/2 = 8L^3 - 3L$ gauge qubits. Since the stabilizer group has dimension $s = 4L^3 + 3L$ and $g+s$ is equal to the total number of physical qubits, there are no logical operators when this code is regarded as a subsystem code.

{\bf\emph{X-Cube Floquet Code:}} We start from the limit where each 4.8.8 layer is initialized in a toric code state.
This can be realized by sequentially measuring yellow checks, blue checks, green checks, and again yellow checks on each layer \cite{Paetznick2022}. We define a period-six measurement sequence, by measuring: 1. yellow checks and on-site checks;  2. blue checks; 3. green checks; 4. yellow checks; 5. blue checks; 6. green checks. Since the yellow check and on-site checks commute, they can be measured simultaneously. This completes our definition of the 3D Floquet code.\footnote{If instead, the code were defined with a shorter measurement schedule (on-site checks, blue, green, yellow) then the code would not return back to X-cube model at any subsequent round and the dimension of the dynamical logical subspace would not be preserved}.


\subsection{Dynamics described by Instantaneous Stabilizer Groups}

\begin{figure}[t]
    \centering
    \includegraphics[width=38mm]{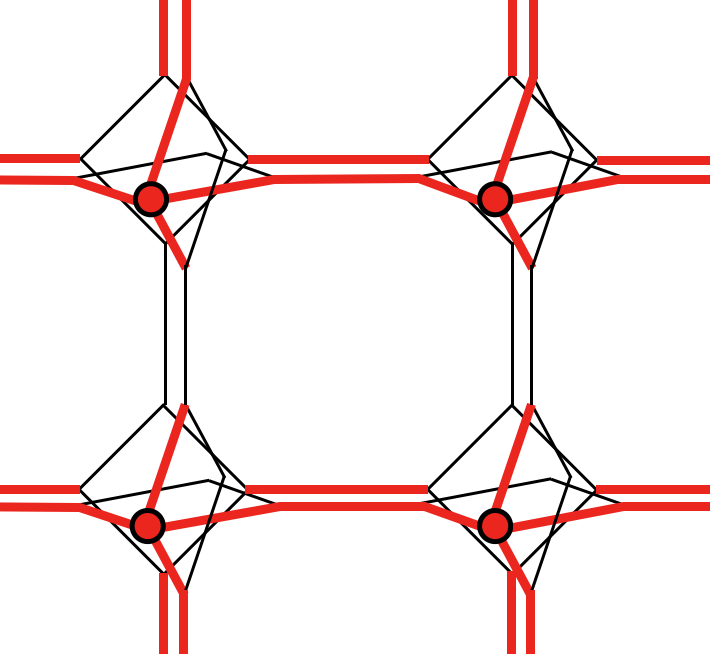}
    \caption{A local snapshot of the non-local membrane stabilizer in the ISG at step 5, given by the product of checks supported on edges and sites highlighted in red. The coloring of the edge (yellow, blue, and green) is not shown here for clarity. There are three such membrane stabilizers in total, one along the $xy$, $yz$, and $xz$ planes.}
    \label{fig3}
    \end{figure}

We shall use the "instantaneous stabilizer group (ISG)" \cite{hastings2021dynamically} formalism to describe the state after each round of measurement. The state under the measurement sequence is stabilized by a Pauli stabilizer group $S$. Whenever we measure a Pauli operator $P$, the operator $\pm P$ (the sign depending on the measurement outcome, chosen according to Born's rule) as one of the generators of a new stabilizer group $S'$. Generators of $S$ that commute with $P$, remain as generators of $S'$. For generators of $S$ that don't commute with $P$, if the product of some of those generators commute with $P$, we consider their product as generators of $S'$. The state we obtain after measuring $P$ is stabilized by the stabilizer group $S'$. 
Suppose we start from the state whose ISG is that of the decoupled-layers of toric code. (This can be realized by starting in a maximally mixed state, and measure yellow, green, blue, and yellow checks sequentially on each layer)
We now identify the evolution of the ISG of our code under the measurement sequence (1. on-site, 2. blue, 3. green, 4. yellow, 5. blue, 6. green, 7. yellow). Here, we separate the yellow check and on-site check measurements into different steps only for the purpose of analyzing the different states of the Floquet code.

\begin{figure*}
    \includegraphics[width=140mm]{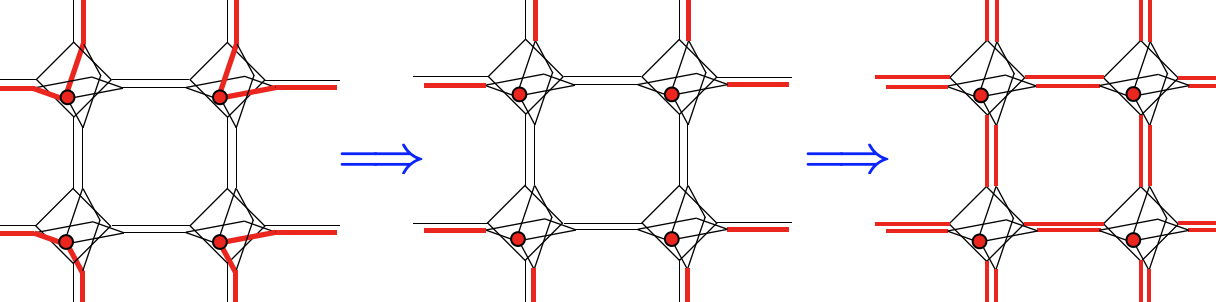}
    \caption{Evolution of the membrane stabilizers from round 6 to round 8 by sequentially measuring blue, green, and yellow checks. Only a local snapshot of each membrane stabilizer is shown. The membrane stabilizer is the product of checks supported on edges and sites highlighted in red. Again, the coloring of the edge(yellow, blue, and green) is not shown for clarity.}
    \label{fig4}
    \end{figure*}
    
\begin{enumerate}
    \setcounter{enumi}{-1}
    \item The ISG is initially generated by yellow checks, yellow plaquette stabilizers, green plaquette stabilizers, and blue plaquette stabilizers. The ISG describes the stabilizer group of $3L$ layers of decoupled toric code. 
    \item After measuring on-site checks, the ISG is generated by on-site checks, yellow checks, yellow plaquette stabilizers, and ``cubic stabilizers", defined as the product of six plaquette stabilizers around each elementary cube. The ISG describes the stabilizer group of the X-cube model, as shown in Appendix \ref{AA}. By a detailed counting of the dimension of the ISG in Appendix \ref{AA}, we can also conclude that the number of encoded logical qubits is equal to that of the X-cube model, which is $6L - 3$.
    \item After measuring blue checks, the ISG is generated by blue checks, yellow plaquette stabilizers, green plaquette stabilizers, cubic stabilizers, and ``zigzag stabilizers". The zigzag stabilizer is defined for every blue plaquette, but it has support on the two layers that are perpendicular to that blue plaquette. The zigzag stabilizer is defined as the product of four yellow checks and four on-site checks, as shown in Fig.\ref{fig2}(a). A detailed counting of the dimension of the ISG after this step is presented in Appendix \ref{AB} and the number of logical operators remain $6L - 3$. The ISG describes the stabilizer group of a model with fracton excitations and the detail is analyzed in Appendix \ref{FTO}.

    \item After measuring green checks, the ISG is generated by green checks, yellow plaquette stabilizers, green plaquette stabilizers, a new set of zigzag stabilizers (See Fig.\ref{fig2}(b)), and cubic stabilizers. The new zigzag stabilizer is similarly defined for every blue plaquette and has support on the two layers that are perpendicular to that blue plaquette. Each new zigzag stabilizer is obtained from the zigzag stabilizer in the previous step by multiplying it by eight blue checks, as shown in Fig.\ref{fig2}(b). The ISG describes the stabilizer group of a model with lineon, planon, and fracton excitations, whose properties are analyzed in Appendix \ref{FTO}. 

    \item After measuring yellow checks, the ISG is generated by yellow checks, yellow plaquette operators, green plaquette stabilizers, blue plaquette stabilizers, and three \emph{non-local} membrane stabilizers living in $xy$, $xz$, and $yz$ plane. A snapshot of the non-local stabilizer around a blue plaquette is shown on \ref{fig3}. The full non-local stabilizer is obtained by taking the product of all such local snapshots over a layer lying in the $xy$, $yz$, or $xz$ plane. Non-local stabilizers on parallel layers are equivalent to each other up to elements in the ISG. \newline
    \indent {We note that if we remove the three non-local stabilizers, the ISG describes $3L$ decoupled layers of toric codes. As discussed in Sec. \ref{sec:dynamics}, the non-local stabilizer in Fig.\ref{fig3} is the product of outer logical operators (logical operators which do not belong to the stabilizer group) living in the blue layers and green layers. The value of outer logical operator on each layer labels the topological sector of the toric code state on that layer. Therefore, the non-local stabilizer in Fig.\ref{fig3} provides a global constraint on the possible topological sectors in the blue and green layers. }\newline

    \item In the next two steps of intra-layer measurements, the ISG at each round is generated by three non-local membrane stabilizers and checks and local stabilizers as that of the 4.8.8 Floquet code. The evolution of the three membrane stabilizers is drawn in Fig.\ref{fig4}. The ISG at each step describes layers of entangled toric codes, which are restricted to certain topological sectors.

    \item If we repeat our measurement schedule and start by measuring on-site checks. The ISG is the same as the ISG we obtain in step 2, which describes the stabilizer group of the X-cube model. Before we measure the on-site checks, the non-local generators of the stabilizer group are products of on-site checks (see Fig.\ref{fig4}). As a result, after measuring the on-site checks, those non-local stabilizers are no longer independent generators of the ISG.
    
\end{enumerate}

\subsection{Dynamics of the outer logical operators}\label{sec:dynamics}
\begin{figure}
    \includegraphics[width=40mm]{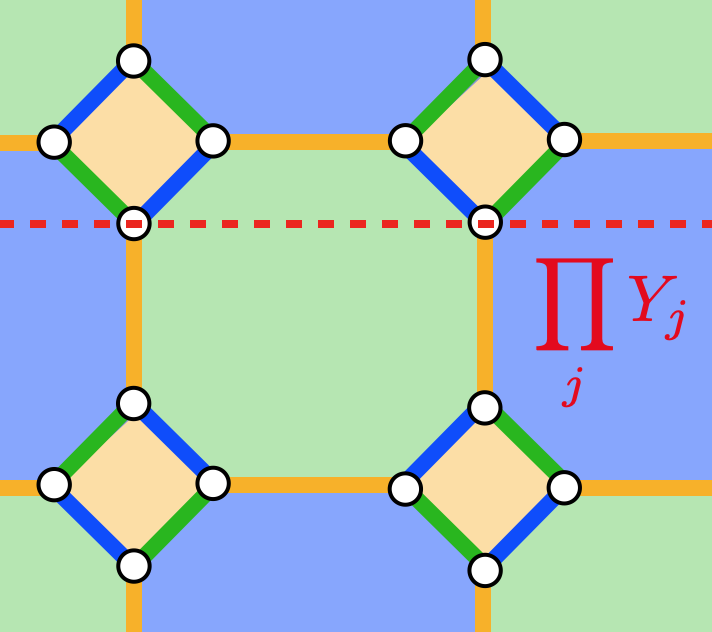} 
    \caption{An outer logical operator defined on each 4.8.8 layer. It is the product Pauli $Y$ along a homologically nontrivial cycle shown in red dashed line.}
    \label{fig6}
\end{figure}

We now discuss the dynamics of the outer logical operators. These logical operators are supported on homologically nontrivial cycles. While these operators commute with all the checks, we cannot infer their values from check measurement outcomes. If we view our code as a subsystem code, then outer logical operators are defined as the operators which do not belong to the stabilizer group.
The number of independent outer logical operators is equal to the number of encoded logical qubits. 
\begin{enumerate}
    \item Initially, we have $3L$ layers of decoupled two-dimensional toric codes. Each layer has size $L\times L$ and gives two outer logical operators (as shown in Fig.\ref{fig6}). In total, there are $6L$ outer logical operators. Let $(\mu, n)$ denote the plane normal to the $\mu$ direction and with $\mu$-coordinate $n$, and let $\gamma^{\nu}_{(\mu, n)}$ denote the path lying on the dual lattice of layer $(\mu, n)$ and passing around the non-contractible loop in the $\nu \neq \mu$ direction. Each outer logical operator is the product of $Y$ along the path $\gamma^{\nu}_{(\mu, n)}$ and we denote it as $S(\gamma^{\nu}_{(\mu, n)})$.

    \item After measuring the on-site checks, the $6L$ outer logical operators remain invariant since they commute with on-site checks. However, they are not independent of each other. For all $\mu \neq \nu$, we have the relation 
    \begin{equation}\label{eq1}
        \prod_{0\leq n \leq L} S(\gamma^{\nu}_{(\mu,n)}) \sim \prod_{0 \leq n \leq L}S(\gamma^{\mu}_{(\nu,n)})
    \end{equation}
    The tilde means the relation holds true up to the multiplication of some on-site checks. We have three pairs of such relations, so the number of independent outer logical operators reduces to $6L - 3$, which is consistent with the number of logical operators in X-cube model. Notice that on-site checks commute with outer logical operators, so outer logical operators remain invariant.\newline
    \indent Another interpretation of equation (\ref{eq1}) is that the measurement outcome of on-site checks destroys three qubits of logical information, so that there are now $6L-3$ qubits of encoded logical information.
    
    \item  Outer logical operators are not measured by intra-layer measurements, so the number of independent outer logical operators remains $6L - 3$ in a Floquet cycle. Outer logical operators on each layer follow the same dynamics as that in the 4.8.8 Floquet code\cite{Paetznick2022}.
    
\end{enumerate}

\section{Error correction properties}\label{sec:errorcorrection}
{The X-Cube Floquet code encodes $6L-3$ logical qubits, and toggles between the X-cube fracton order and weakly-entangled toric code layers.  In this section, we will argue that in the presence of a sufficiently weak rate of measurement errors and single-qubit Pauli errors, that the logical codespace remains robust. }
\subsection{Simplifying error model}
{We consider measurement errors, whereby a check operator is measured and the outcome is incorrectly inferred, along with independent single-qubit Pauli errors. First, a} measurement error is equivalent to a pair of qubit Pauli errors. Consider for example a measurement error for an $XX$ check measurement. This is equivalent to having a $Y$ or $Z$ type error occurring on one of the qubits before measurement, and the same error occurs immediately after measurement. Since a $Y$ or $Z$ type error anti-commutes with the $XX$ check, the $XX$ check measurement outcome is flipped, while no qubit is being acted upon because the same $Y$ or $Z$ type error has occurred twice. Therefore, by reading out the value of syndrome bits, we could apply correction to a measurement error as if it is a pair of Pauli errors.  As a result, it suffices to consider a simplified error model, with perfect check measurements and Pauli errors. {A similar simplification of measurement errors as correlated single-qubit Pauli errors, occurs for the honeycomb code \cite{hastings2021dynamically}.}\newline
\indent Using the fact that $YY$ (yellow) checks commute with on-site checks and on-site checks are always measured right after yellow checks, we can combine the on-site check measurements with the yellow checks measurements. Consider a Pauli error on a qubit. Immediately before the error, we measure a check supported on that qubit, involving some Pauli operator $P_1$ on that qubit. Immediately after, we measure some other check supported on that qubit, involving some Pauli operator $P_2$. We can use these two Pauli operators as a basis for Pauli errors. So there are three types of Pauli errors: $P_1$, $P_2$, and $P_1 P_2$. The $P_2$ error can be commuted with the subsequent check. Then, we can consider a restricted error model similar to that of the honeycomb code \cite{hastings2021dynamically}: Pauli errors may occur on a qubit of a type corresponding to whatever check was measured previously: if an $XX$, $YY$ (or together with on-site check), $ZZ$ check is measured, then subsequently a Pauli error may occur of type $X$, $Y$, $Z$ respectively. \newline

\subsection{Decoding}
Let's consider how to detect errors based on the previously-defined error model. We define our period-six check measurement sequence to be: yellow and on-site checks, blue checks, green checks, yellow checks, blue checks, green checks. The syndrome bits are the yellow, blue, and green plaquette stabilizers. The plaquette stabilizers that can be inferred at each round are: blue, green, yellow, blue, green, yellow. \newline
\indent Suppose after round $r = 1~\text{mod}~6$ a single-qubit error $E~=~X$ occurs. At round $r+1$, exactly one green check is flipped. So the yellow plaquette stabilizer that is inferred by the outcomes at round $r$ and $r + 1$ is flipped. At round $r + 2$, exactly one yellow check is flipped, but the blue plaquette stabilizer inferred at round $r+2$ is not flipped because two checks contained in the blue plaquette stabilizer are flipped. At round $r+3$, no check is flipped, but one green plaquette stabilizer is flipped because one yellow check is flipped at round $r+2$. So, the error $E$ flips two syndrome bits. More generally, a single-qubit error occurred after the measurement at round $r$ flips one check at round $r+1$ and one check at round $r+2$. Therefore, one syndrome bit at round $r+1$ and one syndrome bit at round $r+3$ are flipped. These two flipped syndrome bits nearest neighbors in space, but second nearest in time.\newline
\indent Notice that the error $E$ and the two flipped syndrome bits caused by $E$ are all defined on the same 4.8.8 layer. Therefore, we could construct exactly the same decoding graph on each 4.8.8 layer as that of the 4.8.8 Floquet code. \newline

\indent The zigzag stabilizers (Fig.\ref{fig2}) and non-local stabilizers (Fig.\ref{fig3},\ref{fig4}) can also be used to detect Pauli errors. Suppose a Pauli $Y$ error occurred on some qubit after measuring on-site checks and yellow checks at round $r$. As we previously showed, a blue check which has support on that qubit is flipped and a green plaquette stabilizer is flipped. No zigzag stabilizer is flipped because zigzag stabilizer commutes with Pauli $Y$. At round $r+2$, no plaquette stabilizer is flipped, but the new zigzag stabilizer which contains the flipped blue check in round $r+1$ is flipped. The same argument also applies to non-local stabilizers, where they can detect Pauli errors. Therefore, if we consider zigzag stabilizers and non-local stabilizers appeared in the ISG also as syndrome bits, we could probably design some more efficient decoding algorithm for the fracton Floquet code.

\section{Discussion}
In this paper, we have introduced a 3D Floquet code, which dynamically encodes a sub-extensive number of logical qubits. Our protocol involves two-qubit measurements and has a period-six Floquet period. We have argued that the X-Cube Floquet code has a non-zero error threshold. It would be interesting to design a more efficient decoder which utilizes X-cube stabilizers at step 1 to achieve a better error threshold\cite{brown2020parallelized}. Since there is a Calderbank-Shor-Steane (CSS) version of the 4.8.8 Floquet code \cite{Davydova2022}, it is natural to convert our protocol to a CSS structure, where we only perform type-$X$ or type-$Z$ check measurements. We leave an analysis of the resulting 3D code to future work. Moreover, it would be interesting to investigate what other fracton Floquet codes can be realized with a coupled layer or defect network construction. \newline
\indent Our construction of the 3D Floquet code also implies a 3D lattice Hamiltonian (\ref{eqH}) which only involves 2-body interactions (shown in Appendix \ref{AE}). By tuning the coupling parameters of Hamiltonian (\ref{eqH}), this Hamiltonian can realize phases that represent X-cube fracton topological order and layers of toric code topological order. Those phases match the stabilizers in the ISG at measurement step 1, 4, 5, and 6. At measurement step 2 and 3, we have shown the logical operators of the code exhibit fractonic properties (shown in \ref{FTO}). It remains unclear whether we can find such a fracton topological order in certain parameter regime of Hamiltonian (\ref{eqH}).

 \acknowledgments 
 ZZ, DA, and SV acknowledge enlightening discussions with Jeongwan Haah, Ali Lavasani, and Zhenghan Wang.

\bibliography{collection}

\appendix

\section{Counting the rank of ISG at step 1 }\label{AA}
In this section, we will show that the ISG after measuring on-site checks is equivalent to the stabilizer group of X-cube model.\newline
\indent The ISG is generated by yellow checks, yellow plaquette stabilizers, and cubic stabilizers. The yellow check projects the state of the two qubits down to a two-dimensional subspace, we may regard the pair of qubits living on the same yellow edge as one effective qubit. In this effective qubit space, the lattice reduce to a cubic lattice and each edge contains two effective qubits. The on-site checks further projects the state of the two effective qubits down a two-dimensional subspace, so it is sufficient to define only one effective qubit living on each edge. Then, the yellow plaquette stabilizers act like the vertex stabilizers of X-cube model and the cubic stabilizers act like the cube stabilizers of X-cube model. \newline
\indent By a sanity check, we can count the dimension of the ISG described above. Suppose our coupled-layers have dimension $L\times L \times L$, where $L$ is the number of 4.8.8 layers in each direction. In total, there are $3L$ 4.8.8 layers. On each layer, we have $L^2$ yellow plaquettes, $\frac{1}{2}L^2$ green and blue plaquettes, $2L^2$ yellow, blue, and green edges, and $4L^2$ qubits. There are $L^3$ octahedrons and each octahedron contains $12$ qubits, so the coupled-layers contain $12L^3$ qubits. \newline
\indent Now we can count the number of constraints set by the stabilizers. All the yellow checks give $3L \times 2L^2 = 6L^3$ constraints. On-site checks give $6L^3$ constraints. Yellow plaquette stabilizers give $3L^3$ constraints. Cubic stabilizers give $L^3$ constraints.\newline
\indent However, the above constraints are not all independent and we need to count the number of relations between them. The product of two overlapping yellow checks is equal to the product of on-site checks acting on both ends. This gives $3L^3$ relations. The product of yellow plaquette stabilizers around each octahedron is equal to the product of on-site checks acting on the vertices of that octahedron, which gives $L^3$ relations. On each layer, the product of all yellow checks and yellow plaquette stabilizers is equal to the identity, which gives $3L$ relations. The product of cubic stabilizers along a plane equals to the product of blue and green plaquette stabilizers of two neighboring layers. We can further multiply yellow plaquette operators on those two layers and that should give us the identity. So there are $3L - 3$ relations between cubic stabilizers and yellow plaquette stabilizers. The $-3$ comes from the fact that on each direction the relations for the first $L - 1$ plane tell us the relation for the last one plane.\newline
\indent The number of encoded qubits (logical operators) is given by \#qubits - \#constraints + \#relations = $6L - 3$, which is the number of logical operators in X-cube model defined on an $L\times L \times L$ 3-torus.


\section{Counting the rank of ISG at step 2 and 3} \label{AB}
~~~ In this section we calculate in detail the dimension of the ISG after measuring blue checks at step 2. The counting for step 3 can be done in exactly the same manner.\newline
\indent The ISG is generated by blue checks, yellow plaquette stabilizers, green plaquette stabilizers, zigzag stabilizers(Fig. \ref{fig2}), and cubic stabilizers.\newline
\indent There are $6L^3$ blue checks, $3L^3$ yellow plaquette stabilizers, $\frac{3}{2}L^3$ green plaquette stabilizers, $\frac{3}{2}L^3$ zigzag stabilizers, and $L^3$ cubic stabilizers. This gives in total $13L^3$ constraints.\newline
\indent However, the above constraints are not all independent
and we need to count the number of relations between
them. On each layer, the product of all yellow plaquette stabilizers and green plaquette stabilizers is equal to the product of blue plaquette operators. The product of all blue plaquette operators with blue checks gives the identity. So this gives us $3L$ relations. We also need to consider the relations between cubic stabilizers and yellow plaquette stabilizers as explained in Appendix \ref{AA}. This gives us $3L - 3$ relations. The product of zigzag stabilizers around a cube is equal to the product of yellow plaquette stabilizers or green plaquette stabilizers. This gives us $L^3$ relations.\newline
\indent The number of encoded qubits (logical operators) is given by \#qubits - \#constraints + \#relations = $6L - 3$. Therefore, the number of encoded qubits is preserved.

\section{X-cube Floquet Code as a subsystem code} \label{AC}
We consider an $L\times L \times L$ system with periodic boundary condition. On each 4.8.8 layer, we have $L^2$ yellow plaquettes, $\frac{1}{2}L^2$ green and blue plaquettes, $6L^2$ edges and $4L^2$ qubits. There are $6L^3$ on-site checks. The gauge group is generated by the $24L^3$ check operators. On the torus, not all of the check operators are independent. There are $3L^3$ relations between yellow checks and on-site checks. The product of all yellow, blue, and green checks over a layer is the identity, which gives $3L$ relations. The product of on-site checks around a tetrahedron equals to the product of blue and green checks supported on its edges, which gives $L^3$ relations. Therefore, we determine the gauge group has dimension $20L^3 - 3L$.\newline
\indent The stabilizer group is generated the product of checks on any cycles. The stabilizers corresponding to homologically trivial paths are generated by yellow plaquette stabilizers and cubic stabilizers. The product of check operators on homologically nontrivial cycles are the so-called "inner logical operators"\cite{hastings2021dynamically}.  There are $3L^3$ yellow plaquette stabilizers and $L^3$ cubic stabilizers. At a given measurement round, inner logical operators and outer logical operators act as logical $X$, $Z$ operators. So there are $6L - 3$ inner logical operators as the generators. There are $3L - 3$ relations between cubic stabilizers and yellow plaquette stabilizers as explained in Appendix \ref{AA}. So we find the stabilizer group has dimension $4L^3 + 3L$.\newline
\indent Hence, there are $g = ((20L^3 - 3L) - (4L^3 + 3L))/2 = 8L^3 - 3L$ gauge qubits. Since the stabilizer group has dimension $s = 4L^3 + 3L$ and $g+s$ is equal to the total number of physical qubits, there are no logical operators when this code is regarded as a subsystem code.

\section{Quasiparticles with restricted mobility at step 2 and 3} \label{FTO}
We can study the mobility of quasiparticles of the topological stabilizer code described by ISG at step 2 and 3. Our strategy is to express the stabilizers in an effective Hilbert space, which will greatly simplify the problem. The effective stabilizers at step 3 is equivalent to the effective stabilizers at step 2 up to a basis transformation, so we will focus our analysis on the ISG at step 2. \newline
\begin{figure}[h]
    \centering
    \includegraphics[width=80mm]{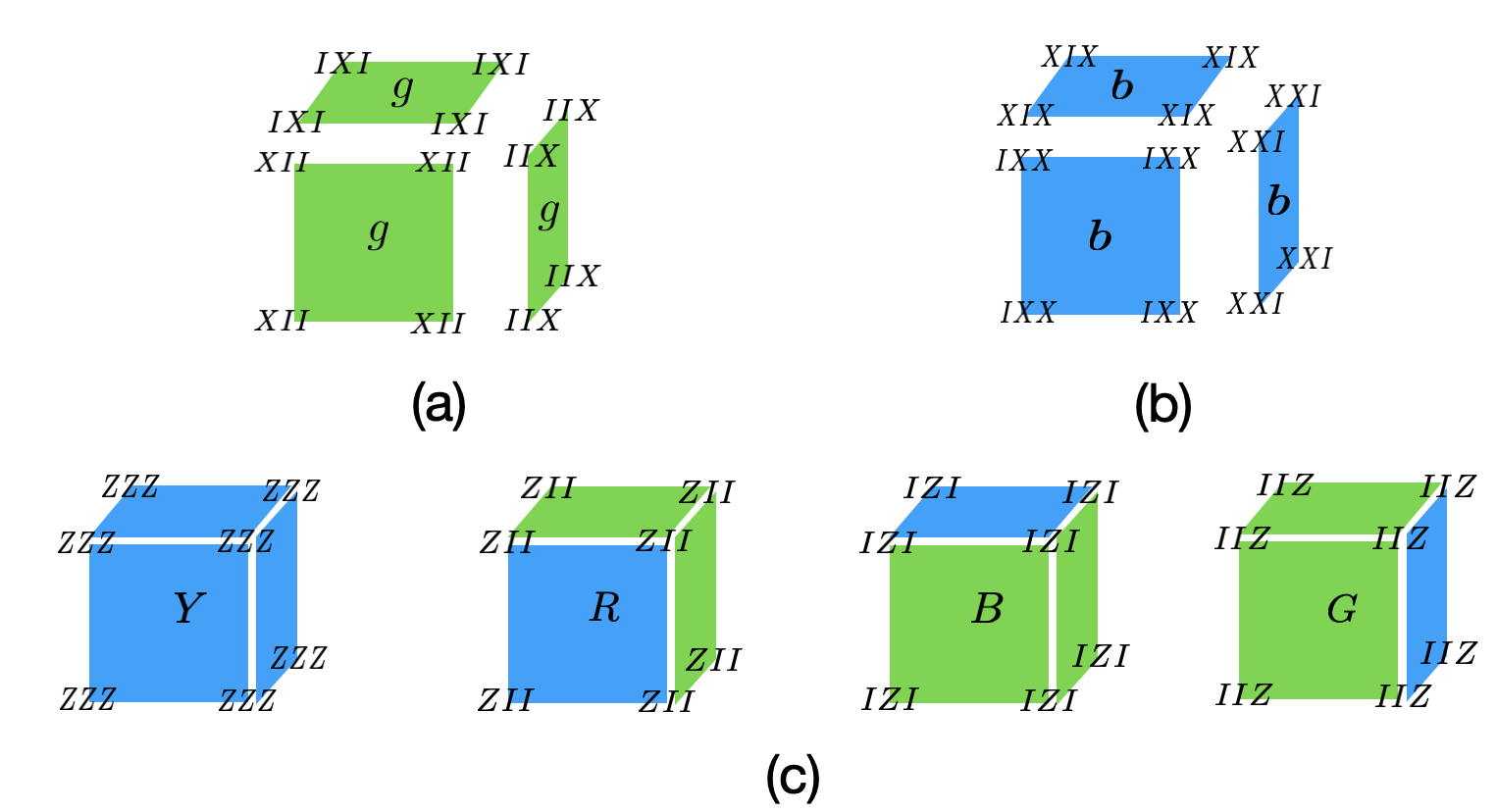}
    \caption{(a): Green plaquette stabilizers in the effective Hilbert space. (b): Blue plaquette stabilizers in the effective Hilbert space. (c): Cubic stabilizers in the effective Hilbert space. Notice that, by our coloring convention, there are only four different types of cubed and they are labeled by $Y, R, B, G$.}
    \label{fig9}
    \end{figure}
\indent The ISG at step 2 is generated by blue checks, yellow plaquette stabilizers, green plaquette stabilizers, zigzag stabilizers, and cubic stabilizers. Since the measurements of blue checks project the two qubits in the support down to a two dimensional subspace, so We can regard that two qubits as one effective qubit. Moreover, in this effective qubit Hilbert space, the yellow plaquette stabilizers act also as two-body operators. Therefore, we can further regard two qubits as one effective qubit. By the above two rounds of simplification, the effective model can be defined on a cubic lattice, where each vertex of the cubic lattice supports three qubits. Each face of the cubic lattice has color green or blue, such that any two neighboring plaquettes on the same 2D plane have different colors. \newline
\begin{figure}[h]
    \centering
    \includegraphics[width=50mm]{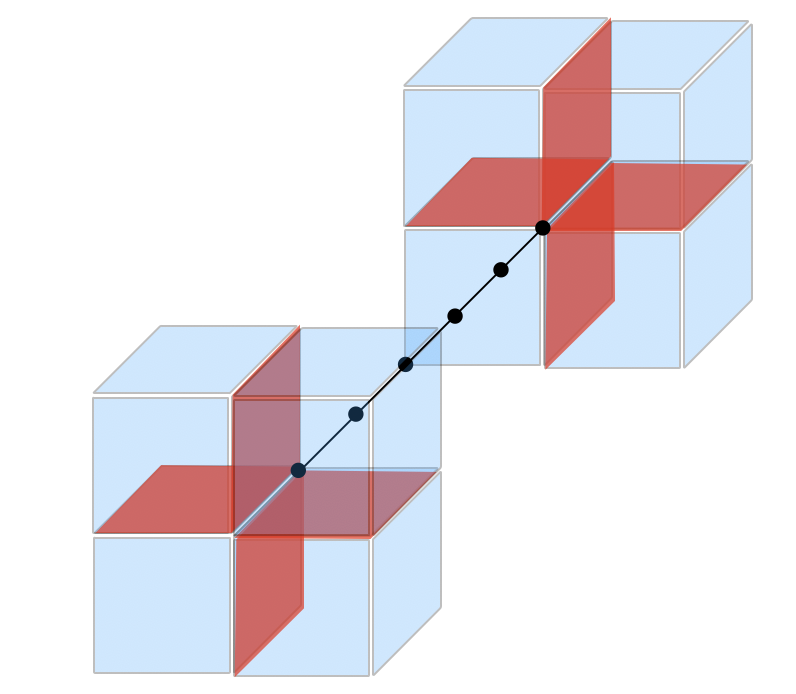}
    \caption{A pair of lineons created by an open string operator. Each lineon corresponds to a violation of two green plaquette stabilizers and two zigzag stabilizers, which are shown in red color. Each black dot in the string operator represents the operator $IZZ$.}
    \label{fig10}
    \end{figure}
\indent Now we describe the stabilizers defined on the effective cubic lattice (See Fig.\ref{fig9}. for their definitions). The green plaquette stabilizers are defined similarly on green faces. Depending on the direction of the green face, it acts on different qubits on each vertex. The zigzag stabilizers are defined on blue faces, and the qubits in support depend on the direction of that blue face. The cubic stabilizer is defined for each cube in the cubic lattice. Since green plaquette stabilizer is already included in the stabilizer group, we can define the cubic stabilizer as the product of blue plaquette operators around a cube. We notice that green plaquette stabilizer and zigzag stabilizer only consist Pauli $X$ operators and the cubic stabilizer only consists Pauli $Z$ operators.\newline
\begin{figure}[h!]
    \centering
    \includegraphics[width=50mm]{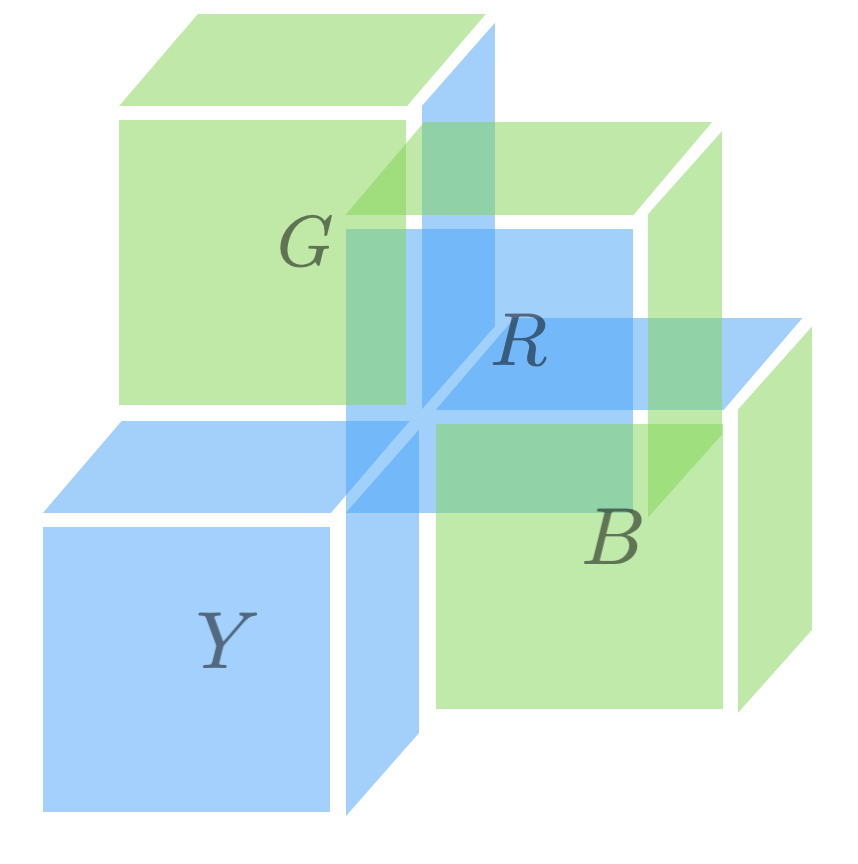}
    \caption{A unit cell for the A sublattice in the effective model. The color of each plaquette and the type of each cube are shown explicitly.}
    \label{fig11}
    \end{figure}
\indent {\bf \emph{Lineon excitations:}} A lineon excitation corresponds to violation
of two green plaquette stabilizers and two zigzag stabilizers (See Fig.(\ref{fig10})). There are three types of lineon and we will label them as $l_x$, $l_y$, and $l_z$ based on their mobility direction. A pair of lineons of the same type can be created at the endpoints of a string operator(See Fig.\ref{fig10}). The lineons obey a triple fusion rule, such that $l_x \times l_y \times l_z = 1$. Two adjacent lineons move like a planon. For example, two neigboring  $x$-type lineons separated in the $y$-direction can move in the $xz$ plane.  \newline
\begin{figure}[ht]
    \centering
    \includegraphics[width=80mm]{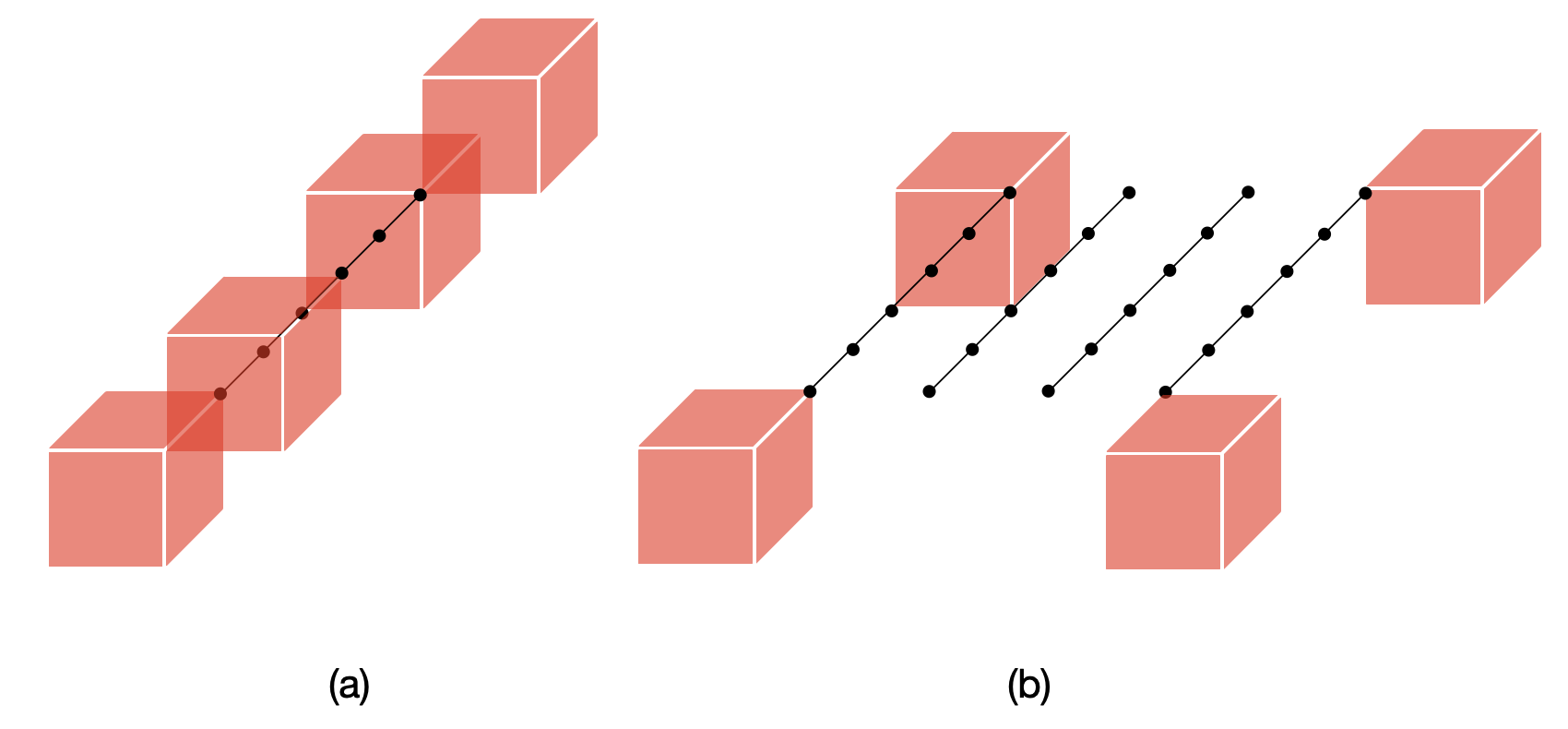}
    \caption{(a):A pair of $Y$ type fracton and $R$ type fracton bound states are created at the endpoints of a string operator. Each black dot in the string operator represents the operator $XII$. (b): A membrane operator which creates four fractons at the corner. Each black dot in the membrane operator represents $XII$ and the four fractons are $f_{Y}^{A}$.}
    \label{fig12}
    \end{figure}
\indent {\bf \emph{Fracton excitations:}} In the cubic lattice, we can divide all the cubes into $A$ and $B$ sublattices. We view four cubes sharing a same vertex on the same sublattice as a unit cell of the system (See Fig.\ref{fig11}). We further divide cubes in the each sublattice into four different types labeled by $Y, R, B, G$ (This is the same labeling as in Fig.\ref{fig9}). Therefore, a violation of a cube stabilizer can be labeled by $f^{\alpha}_{\beta}$, where the upper index $\alpha \in \{A, B\}$ labels the sublattice of that cube and the lower index $\beta \in \{Y, R, B, G\}$ labels the type of that cube. \newline
Each individual $f^{\alpha}_{\beta}$ is a fracton and is immobile. Four of them can be created at the corners of a membrane operator (See Fig.\ref{fig12}(b)). The fractons obey the fusion rule $f_{Y}^{A} \times f_{\beta}^{A} \times f_{Y}^{B} \times f_{\beta}^{B} = 1$, where $\beta \in \{R, B, G\}$. The bound state of a $Y $ type fracton and a $R$, or $G$, or $B$ type fracton on different sublattice moves like a planon. The bound state of a $Y$ type fracton and a $R$, or $G$, or $B$ type fracton on the same sublattice moves like a lineon in different direction (See Fig.\ref{fig12}(a)). Such a lineon cannot locally decomposed into two planons. Three lineons moving in different directions don't fuse into identity(See Fig.\ref{fig13}). The above fracton excitations are summarized in Table \ref{table:1}.\newline
    
    \begin{figure}[ht]
    \centering
    \includegraphics[width=50mm]{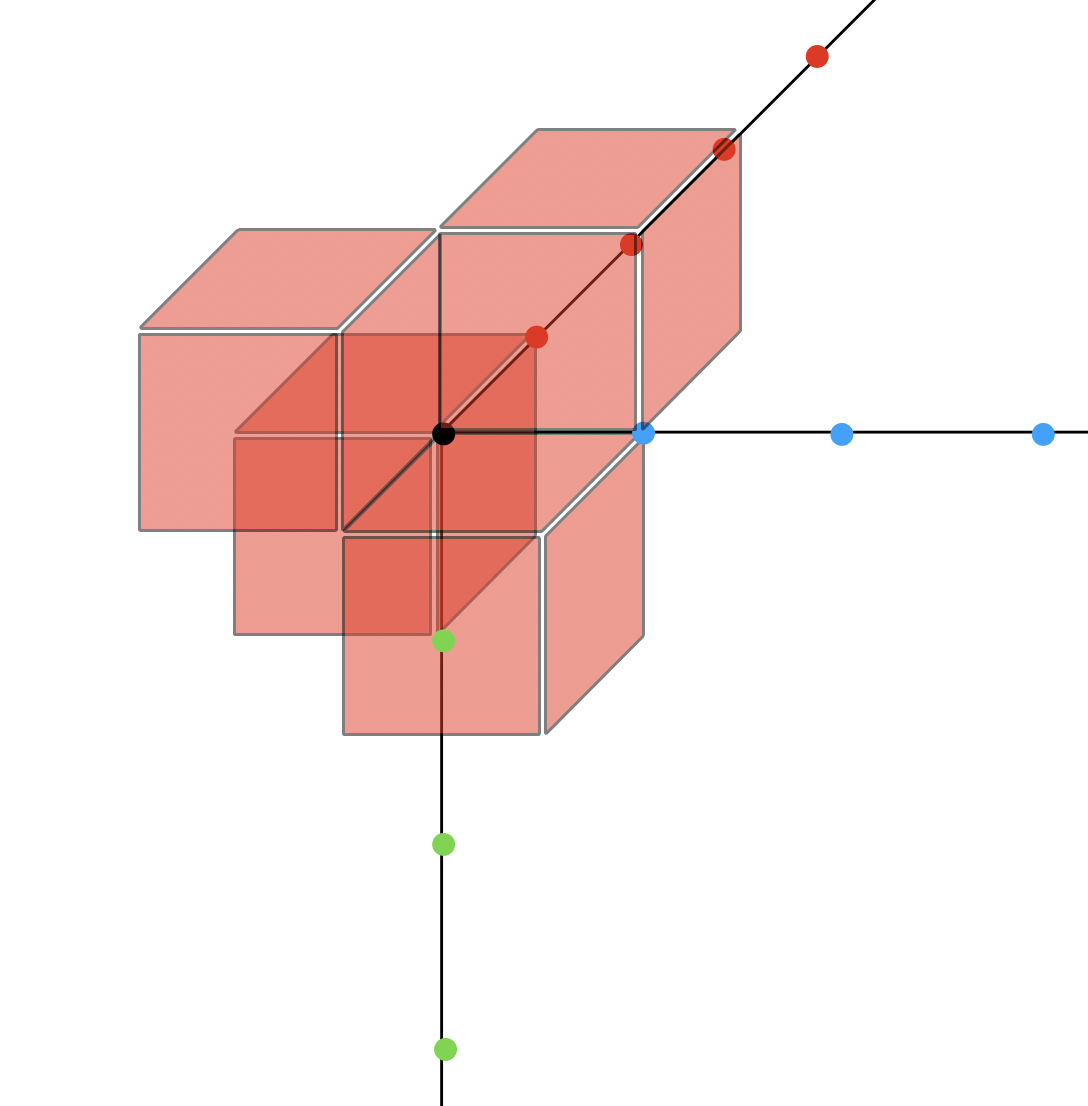}
    \caption{Three lineons (bound states of two fractons in the same sublattice) don't fuse into the identity when they meet at a corner. Four fractons drawn in red are created at the corner. The line of red dots is a string of $XII$, which creates a pair of $x$ direction lineons. The line of blue dots is a string of $IIX$, which creates a pair of $y$ direction lineons. The green line is a string of $IXI$, which creates a pair of $z$ direction lineons. The black dot represents the operator $XXX$.}
    \label{fig13}
    \end{figure}


\section{A 3D Lattice model}\label{AE}
We could write down a Hamiltonian which involves only two body interactions on the 3D coupled-layer lattice in Fig.\ref{fig:488}.
\begin{equation}\label{eqH}
\begin{split}
    H = &-J_y \sum_{<ij> \in \text{yellow}}\sigma_{i}^{y}\sigma_{j}^y -J_b \sum_{<ij> \in \text{blue}} \sigma_{i}^{\alpha_{ij}}\sigma_{j}^{\alpha_{ij}}\\ &- J_g \sum_{<ij> \in \text{green}}\sigma_{i}^{\alpha_{ij}}\sigma_{j}^{\alpha_{ij}}
    -h\sum_{(i,j) \in v}\sigma_{i}^y \sigma_{j}^y
\end{split}
\end{equation}
The first three terms describe decoupled layers of Kitave model\cite{Kitaev:2005hzj} defined on 4.8.8 lattice. Notice that here we use the color of each edge to group each interaction. $\alpha_{ij} \in \{x, y, z\}$ labels the type of each edge. So for a blue or green edge, the operator supported on that edge can be either $\sigma_{i}^z \sigma_{j}^z$ or $\sigma_{i}^x \sigma_{j}^x$ depending on the direction of that edge $<ij>$. For a yellow edge, the operator supported on that edge is always $\sigma_{i}^y \sigma_{j}^y$. The last term couples different layers together and acts on the two sites $(i, j)$ living on the same vertex $v$. \newline
\indent Measuring, for example, yellow checks in the fracton Floquet code can be thought of as tuning the yellow checks interaction to infinity. Therefore, the state of the Floquet code obtained after measuring certain checks can be interpreted as the ground state of Hamiltonian (\ref{eqH}) in certain parameter regime. 
\begin{enumerate}
    \item We first set $h = 0$ and $J_y \gg J_b, J_g$. This corresponds to the decoupled toric code limit of Hamiltonian (\ref{eqH}). We increase $h$ while keeping $J_y,~J_b,~\text{and}~J_g$ fixed. When $h \gg J_b, J_g$, the Hamiltonian (\ref{eqH}) matches exactly the isotropic layer construction of X-cube model in ref \cite{vijay2017isotropic}. The inter-layering coupling $\sigma_{i}^y \sigma_{j}^y$ condensed composite flux loops in decoupled toric code layers, which then leads to the X-cube model. This parameter regime corresponds to measuring on-site checks at step 1 in the X-cube Floquet code.
    
    
    
    \item Measuring yellow checks at step 4 corresponds to taking $J_y \gg J_b, J_g \gg h$ of Hamiltonian (\ref{eqH}). We first set $h = 0$. The effective Hamiltonian given by perturbation theory is simply decoupled layers of Toric code. Since we know Toric code is stable against local perturbations, if we perturb decoupled-layers of Toric code with on-site $YY$ operators, the long distance physics is unaffected. This matches the ISG at step 4, which contains Toric code stabilizers as the only local stabilizers. The above result also applies to step 5 and 6.
\end{enumerate}
~~\newline
\begin{table}[h]
        \centering
    \begin{tabularx}{0.45\textwidth} { 
  | >{\raggedright\arraybackslash}X 
  | >{\centering\arraybackslash}X 
  | >{\raggedleft\arraybackslash}X | }
 \hline
 Quasiparticles & Type \\
 \hline
 $f^{\alpha}_{\beta}$,~ $\alpha$ $\in \{ A,B\}$,~ $\beta$ $\in \{Y, R, B, G\}$  & Fracton   \\
\hline
$f^{A(B)}_{Y}$ $\times$ $f^{A(B)}_{R}$ & $x$-direction lineon \\
\hline
$f^{A(B)}_{Y}$ $\times$ $f^{A(B)}_{B}$ & $z$-direction lineon \\
\hline
$f^{A(B)}_{Y}$ $\times$ $f^{A(B)}_{G}$ & $y$-direction lineon \\
\hline
$f^{A(B)}_{Y} \times f^{B(A)}_{R}$ & $yz$ planon \\
\hline
$f^{A(B)}_{Y} \times f^{B(A)}_{B}$ & $xy$ planon \\
\hline
$f^{A(B)}_{Y} \times f^{B(A)}_{G}$ & $xz$ planon \\
\hline
\end{tabularx}
\caption{Quasiparticles created by a violation of cube stabilizers.}
\label{table:1}
\end{table}
\section{Relations between ISG at step 2 and step 5}
In the honeycomb code, there exists an automorphism between ISG at round $r$ and round $r+3$. The logical operators undergo a "magnetic $\leftrightarrow$ electric" transitions\cite{hastings2021dynamically}. In our code, we obtain a more general group homomorphism between ISG at step 2 and round 5, and also between ISG at round 3 and round 6. We shall use the ISG at round 2 and round 5 to illustrate this idea.\newline
\indent We label ISG at round 2 as $S(2)$ and ISG at round 5 as $S(5)$. Their common subgroup is generated by blue checks, yellow plaquette stabilizers, green plaquette stabilizers, green plaquette stabilizers, and cubic stabilizers. We label this common subgroup as $S_c$. Since $S(3)$ and $S(6)$ has the same rank, there exists a unitary operator $U$ such that $US(3) U^{-1} = S(6)$. However, it is unclear how $S_c$ is transformed under $U$. There is a potential in studying whether there exists a group automorphism between $S_c$ at step 2 and step 5.

\end{document}